\documentclass[sigconf]{acmart}
\usepackage{subcaption}
\usepackage[normalem]{ulem}
\usepackage{xfrac}
\usepackage{siunitx}
\usepackage{float}
\usepackage{siunitx}
\usepackage{tabularx, booktabs, makecell, caption}
\usepackage{multirow}
\usepackage{array}
\newcolumntype{M}[1]{>{\centering\arraybackslash}m{#1}}

\AtBeginDocument{%
  \providecommand\BibTeX{{%
    \normalfont B\kern-0.5em{\scshape i\kern-0.25em b}\kern-0.8em\TeX}}}

\copyrightyear{2020}
\acmYear{2020}
\setcopyright{acmlicensed}\acmConference[ICAIF '20]{ACM International Conference on AI in Finance}{October 15--16, 2020}{New York, NY, USA}
\acmBooktitle{ACM International Conference on AI in Finance (ICAIF '20), October 15--16, 2020, New York, NY, USA}
\acmPrice{15.00}
\acmDOI{10.1145/3383455.3422521}
\acmISBN{978-1-4503-7584-9/20/10}

\begin{document}

\title{A Hybrid Learning Approach to Detecting Regime Switches in Financial Markets} 

\author{Peter Akioyamen}
\email{pakioyam@uwo.ca}
\affiliation{%
  \institution{Western University}
  \city{London}
  \state{Ontario}
  \country{Canada}
}

\author{Yi Zhou Tang}
\email{ytang294@uwo.ca}
\affiliation{%
  \institution{Western University}
  \city{London}
  \state{Ontario}
  \country{Canada}  
}

\author{Hussien Hussien}
\email{hhussien@uwo.ca}
\affiliation{%
  \institution{Western University}
  \city{London}
  \state{Ontario}
  \country{Canada}
}

\renewcommand{\shortauthors}{Akioyamen et al.}

\begin{abstract}
  Financial markets are of much interest to researchers due to their dynamic and stochastic nature. With their relations to world populations, global economies and asset valuations, understanding, identifying and forecasting trends and regimes are highly important. Attempts have been made to forecast market trends by employing machine learning methodologies, while statistical techniques have been the primary methods used in developing market regime switching models used for trading and hedging. In this paper we present a novel framework for the detection of regime switches within the US financial markets. Principal component analysis is applied for dimensionality reduction and the $k$-means algorithm is used as a clustering technique. Using a combination of cluster analysis and classification, we identify regimes in financial markets based on publicly available economic data. We display the efficacy of the framework by constructing and assessing the performance of two trading strategies based on detected regimes.
\end{abstract}


\maketitle

\section{Introduction}
\label{section:intro}
There is a variety of empirical evidence and corresponding studies which suggest that the time-series behaviors of both economic and financial data may not express invariable patterns through time. Intervals of divergent behavior in financial and economic time-series are often referred to as regimes. The relations between financial markets, global economies, asset valuations, and world populations has made the detection of market regime switches of importance to researchers and practitioners. The financial markets may be qualitatively characterized by various non-disjoint regimes. Periods of high and low volatility influence the actions of traders and investors alike. Often, economic expansions and contractions dictate the long-term market environment, while certain assets can exhibit mean-reverting or trending behaviors in the short-term \cite{chaudhuri2003mean, bali2008nonlinear, corhay1993common}. The presence of herding, originally studied in \cite{banerjee1992simple}, as well as contrarianism may generate changes in market conditions such as misalignments in price action and fundamental asset valuations, among others \cite{cipriani2008herd, chiang2010empirical, tan2008herding, yan2012industry, park2011herding}. The aforementioned behaviors and properties which present themselves over time characterize unique sets of market regimes which may occur simultaneously.
    
To capture the dynamic behavior of such time-series, it is often preferable to employ multiple models. Particularly, when attempting to model financial time-series over a set of regimes, the Markov switching model \cite{hamilton1989new}, also known as the regime switching model, has been studied and applied. Generally, the Markov switching model is governed by an unobservable state variable and time-varying transition probabilities. Despite the many empirical applications and theoretical developments involving Markov switching models, a necessary condition for the method is the need to know the number of regimes which exist prior to estimating model coefficients. This limits the flexibility of the regime switching model in two contexts. The first being applications where modeling and forecasting the time-series may be undesirable, and identifying the occurrence of the underlying regimes may be all that is sought after. The second consists of contexts where it may be beneficial to directly infer the number of regimes from data.

In this study, we seek to identify meaningful regimes financial time-series may be subject to at different points in time without explicitly modeling the time-series as an autoregressive process. As such, we provide an alternative approach to detecting regime switches through the use of a hybrid learning framework. The $k$-means clustering algorithm is used to detect points in time which share underlying characteristics, grouping the data into distinct regimes. We show how clustering may be a desirable approach to regime detection as the number of regimes in the data can be passively inferred. Classification models are fit to the regimes detected by $k$-means, and are then used to predict the regime corresponding to each observation in the out-of-sample data. Based on the historical performance of assets during regimes identified in training data, trading strategies are constructed, displaying one application of the approach in trading and portfolio management. 
    
The paper is organized as follows. Related work is presented in Section~\ref{section:lit}. In Section~\ref{section:data} we discuss the data set used in this work and the related pre-processing techniques applied to the data, including the use of principal component analysis. Section~\ref{section:rd} outlines the framework for regime detection using unsupervised learning. Specifically, the combination of cluster analysis and classification used in the unsupervised learning approach along with in-sample model performance are presented in Sections~\ref{section:rdcluster} and ~\ref{section:rdclass}. In Section~\ref{section:results} we display the efficacy of this approach and a potential use-case of this framework. The construction and performance of two trading strategies based on the detection of regime switches is discussed. The paper is then concluded in Section~\ref{section:conclusion}.

\section{Literature Review}
\label{section:lit}
The Markov switching model, which governs shifts in the coefficients of an autoregression through a discrete-state Markov process, was originally presented in \cite{hamilton1989new}. Extensions of this work were quickly established with the use of the EM algorithm for maximum likelihood estimation along with the application of the cointegrated vector autoregressive process subject to Markovian shifts \cite{hamilton1990analysis, krolzig1996statistical}. The Markov switching model of conditional mean was extended to incorporate the switching mechanism into conditional variance models as well. The autoregressive conditional heteroscedasticity (ARCH) and generalized autoregressive conditional heteroscedasticity (GARCH) models were studied with Markov switching in \cite{cai1994markov, hamilton1994autoregressive}, among others. GARCH with Markov switching was used to model and forecast price volatility in gold; it was found that trading gold futures based on this model resulted in higher cumulative return compared to other GARCH type models \cite{sopipan2012forecasting}. The regime switching ARCH model is also seen in the modeling of Taiwanese stock market volatility \cite{chen2000switching}. The Markov switching model and its variants have been applied widely in the analysis of economic and financial time-series. The analysis and forecasting of economic business-cycles can be seen in \cite{goodwin1993business}, which has gone on to be an important application of the Markov switching model. Among other use-cases, variants of the Markov switching model have been employed to analyze the behavior of interest rates and foreign exchange rates as well \cite{gray1996modeling, garcia1996analysis, engel1990long}. A multivariate extension of the regime switching model is used in \cite{sarno2000cost} where regime dependence is found in the relationship between the stock index spot and futures markets. The model has seen applications in trading and portfolio management for hedging as well as price and volatility modeling \cite{alizadeh2008markov, bock2009regime, chen2013markov, ma2011portfolio, fong2002markov, dai2010trend, mount2006predicting, hardy2001regime}. Naturally, many of these applied works are concerned with modeling the time-series and its behaviors rather than the identification of the regime switches and persistence themselves.

To the best of our knowledge, there has been no attempt to apply a hybrid approach to regime detection which uses both unsupervised and supervised learning algorithms. A related approach has been taken in \cite{zhong2017comprehensive} where authors seek to forecast the daily directional movements of the SPDR S\&P 500 ETF closing price. Beyond this, \cite{liao2008mining, nanda2010clustering} apply clustering as a technique for data mining in the stock market, assisting in the extraction of ancillary investment information or directly in the construction of investment portfolios. The study and application of unsupervised learning for regime switch detection is extremely limited, and this work attempts to fill this gap.

\section{Data and Methods}
\label{section:data}
In this study, open-source Federal Reserve Economic Data (FRED) from aggregator Quandl is used \cite{quandl}. The data used ranges from January 7, 1994 to April 1, 2020, with each unique time-series tracking an economic statistic by its day-on-day percentage change. These time-series can be grouped into eight categories: growth statistics, price and inflation indices, money supply statistics, interest rates, employment statistics, income and expenditure statistics, debt levels, and miscellaneous indicators which report housing starts and gross private domestic investment, among others. In practice, the release of economic data can be infrequent and impacts the way both investors and traders participate in the markets. Projections are often made in between data releases; although useful, these expectations are often given context using the latest official data releases. Different data sources and aggregators can exhibit inconsistencies in projections and forecasts depending on the economic statistic in question. Consequently, in this study, missing values which naturally occur in between reporting dates of a statistic were imputed using the latest observed percentage change in the data. This coincides with the most recent official information a trader or investor may have at a given point in time. Some time-series observations and economic statistics were removed from the data set prior to analysis due to having a late first reporting date or an extremely low frequency of reporting. The resulting matrix contains 48 time-series (columns) which track various macroeconomic statistics within the USA over 6625 time observations (rows). The data is split into in-sample (training) and out-of-sample (test) sets. The training data consists of all observations up to and including December 31, 2013 (5051 observations). Testing data contains observations from January 1, 2014 onward (1574 observations). 

\subsection{Principal Component Analysis}
\label{section:princomp}
Let $\mathbf{R}$ be the data matrix of size $T \times S$, where $T$ is the total number of time observations considered, and $S$ is the total number of economic statistics used as features in this work. Every entry $r_{ij}$ of matrix $\mathbf{R}$ represents the day-on-day percentage change of economic statistic $j$ at time $i$. Principal component analysis (PCA) is performed on the matrix $\mathbf{R}^{*}$, the column-standardized version of $\mathbf{R}$, using its singular value decomposition (SVD). The resulting matrix, denoted $\mathbf{X}$, contains 48 principal components. The derivation of PCA and its properties are discussed in \cite{hotelling1933analysis, Jolliffe, mardia1979multivariate, jolliffe2016principal}. 
    
Principal component analysis is used as a dimensionality reduction and feature extraction technique. We seek to capture maximal variance from the original data set using linear combinations of the economic statistics at each point in time. The resulting principal components are uncorrelated time-series observations. This allows for a lower dimensional representation of the original data; we elect to use 26 principal components in the analysis which collectively contain $90\%$ of the variance present in the original data. Table~\ref{tab:PCA} displays a summary of the results produced by PCA -- the eigenvalue of each dimension represents the amount of variance the dimension captures.

\begin{table}[h] 
\centering
\caption{Results of principal component analysis.}
\label{tab:PCA}
\begin{tabular}{lccc}
\hline
Principal Component & Dim 1 & Dim 2 & Dim 3 \\
\hline
Eigenvalue & 8.987 & 3.089 & 3.023 \\
\% of variance & 18.72 & 6.44 & 6.30 \\
Cumulative \% of variance & 18.72 & 25.16 & 31.46 \\ 
\hline
\hline
Principal Component & Dim 16 & Dim 32 & Dim 48  \\
\hline
Eigenvalue & 1.021 & 0.356 & \num{9.284e-7} \\
\% of variance & 2.12 & 0.74 & 0.00 \\
Cumulative \% of variance & 73.94 & 96.76 & 100.00 \\ 
\hline
\end{tabular}
\end{table}

The loadings of the first two principal components allow for their characterization with respect to the economic statistics contained in the original data. The first principal component primarily captures economic productivity within the USA. The top four factors contributing to the first principal component include \textit{gross domestic product}, \textit{real gross domestic product}, \textit{all employees: total nonfarm payrolls}, and \textit{federal debt: total public debt as percent of gross domestic product}. The top four contributors to the second principal component include \textit{personal saving rate}, \textit{real disposable personal income}, \textit{disposable income}, and \textit{personal consumption expenditures}. As such, principal component two primarily represents economic income and expenditure-related factors.

\section{Regime Detection}
\label{section:rd}
In this regime detection framework, we perform a cluster analysis on the principal components to identify intervals in which the time-series exhibit similar underlying behaviors and characteristics. Through this we are able to find distinct groupings in which each observation may belong, and these groupings are the regimes that are detected in the data. The process of regime detection on the out-of-sample principal component data is framed as a classification task where the regimes detected through cluster analysis act as the classification labels for the in-sample data. Although clustering algorithms such as $k$-means and $k$-medoids may be extended to cluster out-of-sample data by assigning each of the new out-of-sample observations to the pre-existing centroid or medoid with the smallest distance or dissimilarity, the ability to cluster out-of-sample data is not necessarily generalizeable to other algorithms. By combining cluster analysis and classification in this way, we are able to develop a framework that is model agnostic with respect to the clustering algorithm chosen in a given instance.

\subsection{Regime Clustering}
\label{section:rdcluster}
\begin{figure}[h]
    \centering
    \includegraphics[width=\linewidth]{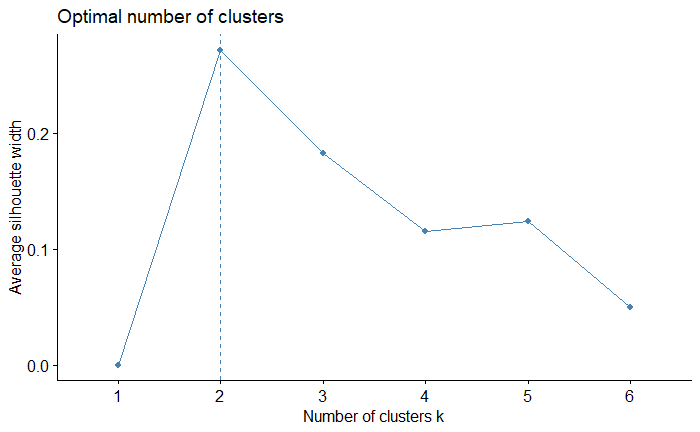}
    \caption{Average silhouette width across 6 values of $k$.} 
    \label{fig:silhouette}
\end{figure}

We perform cluster analysis on the training data set using the $k$-means clustering algorithm with euclidean distance \cite{hartigan1979algorithm}. The value of $k$ used in this work is inferred from the training data through the average silhouette width method \cite{rousseeuw1987silhouettes, kaufman2009finding}. The selection of $k$ is automated by choosing the value of $k$ which maximizes the average silhouette width. In this work we restrict the values of $k$ considered to be in the range $1$ to $6$. The average silhouette width for the possible values of $k$ considered is displayed in Figure~\ref{fig:silhouette}. From this we select $k=2$ such that the training data is segmented into two detected regimes. The $k$-means algorithm is run with $100$ random initializations to ensure the stability of the clusters.

\begin{figure*}[h]
  \begin{subfigure}{0.49\textwidth}
    \includegraphics[width=\linewidth]{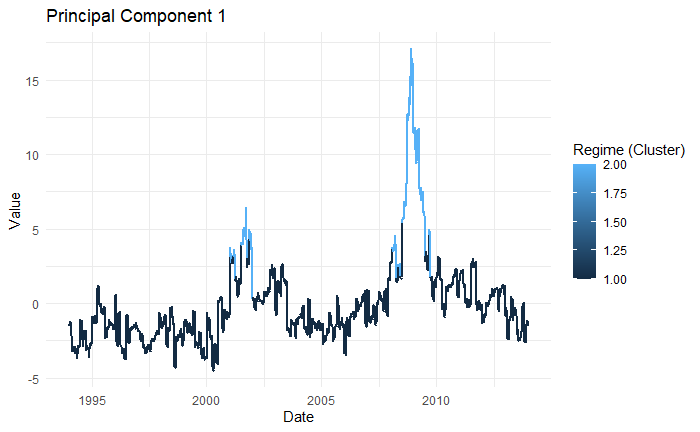}
    \subcaption{Principal component 1 colored by regime after clustering.}\label{fig:PC1_regimes}
  \end{subfigure}
	\hfill
  \begin{subfigure}{0.49\textwidth}
    \includegraphics[width=\linewidth]{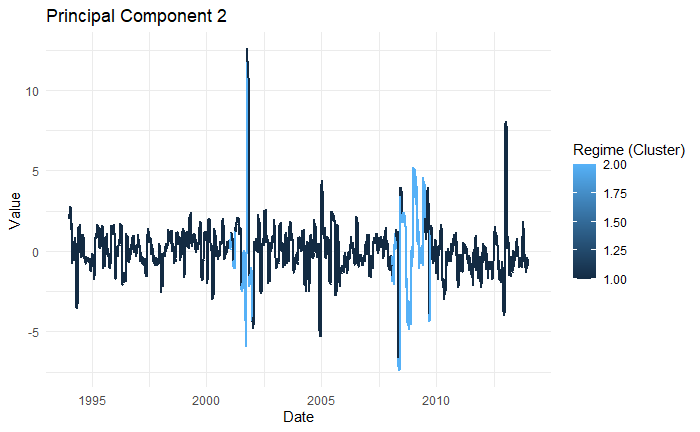}
    \subcaption{Principal component 2 colored by regime after clustering.}\label{fig:PC2_regimes}
  \end{subfigure}
\caption{The first two principal components colored according to the regimes after performing clustering on the in-sample time period (1994--2013).}
\label{fig:PCs_regimes}
\end{figure*}

Figure~\ref{fig:PCs_regimes} displays the first two principal components colored by the occurrence of regimes 1 and 2. It can be seen that meaningful and distinct regimes are detected from the data. Both the latter half of the 2000--2002 dot-com bubble as well as the majority of the 2007--2009 financial crisis are captured in regime 2 (colored light blue). This is indicative of regime 2 coinciding with economic crises and market bubbles originating from the US financial markets. Regime 1 (colored black) coincides with non-crisis periods where the economy and financial markets within the US are generally operating in the absence of abnormal market behavior and events. 

\subsection{Regime Classification}
\label{section:rdclass}
Supervised learning models are fit to the in-sample principal component data and regimes, and is subsequently used to predict the regimes of the out-of-sample observations. The models considered include linear discriminant analysis (LDA), quadratic discriminant analysis (QDA), logistic regression, decision tree classifier, adaptive boosting (AdaBoost), and naive bayes classifier. Each of the models used in this work take the first 26 principal components as input, which may be denoted $X_{1},\dots,X_{26}$, and are trained over the time period 1994--2013. The models produce a binary output which is then mapped to a value in $\{1,2\}$, representing the regime predicted. Model performance is assessed using 10-fold cross-validation. Table~\ref{tab:OOO} displays the in-sample performance results including area under the ROC curve (AUC), accuracy, and F1 score for each model.

\begin{table}[ht]
    \centering
    \caption{Model performance on in-sample data.}
    \label{tab:OOO}
    \begin{tabular}{lccc}
    \hline
    Model & AUC & Accuracy & F1 Score \\
    \hline
    LDA & 0.9996 & 0.9909 & 0.9950 \\
    QDA & 0.9912 & 0.9661 & 0.9813 \\
    Logistic Regression & 1.0000 & 0.9996 & 0.9998 \\
    Decision Tree & 0.9957 & 0.9990 & 0.9995 \\
    AdaBoost & 1.0000 & 0.9994 & 0.9997 \\
    Naive Bayes & 0.9852 & 0.9596 & 0.9777 \\
    \hline
    \end{tabular}
\end{table}

All models demonstrate a strong ability to classify regimes, with each model achieving an accuracy above 0.95 and a minimum AUC of 0.98. The naive bayes classifier performs the poorest across all three metrics while both logistic regression and AdaBoost produce ideal ROC curves with an AUC of 1. These performance metrics may assist in model selection for strategy construction by assessing each model's ability to generalize on the out-of-sample data. Despite this, we employ all models in constructing unique trading strategies to explore the existence of any relationships between the performance evaluation metrics and performance of the trading strategies considered.

\section{Backtest Results}
\label{section:results}
We display the efficacy of this hybrid approach through the construction of two trading strategies, tail-hedging and tactical allocation. The decision rules governing the assets in a specific strategy are defined by the historical performance of the assets through the regimes identified in the in-sample data. The tail-hedging strategy considered here is executed for a single asset, where performance is benchmarked against the underlying asset's passive buy-hold performance (Table~\ref{tab:passive}). Tactical allocation is a rotational strategy which consists of multiple assets. The performance of the strategy is benchmarked against a passive buy-hold strategy applied to the S\&P 500. Every asset is traded through its respective continuous front month futures contract. The performance of each strategy is assessed over the out-of-sample time period, 2014--2020.

The trading mechanism for both strategies is as follows: at the end of day $t$, new data is released and pre-processed giving inputs $X_{1},\dots,X_{26}$. These are used by the selected classification model to predict the current regime on day $t+1$. Based on the predicted regime, an appropriate trade is executed on day $t+1$ in accordance with a strategy's decision rules. Positions are entered at the start of day $t+1$ and held until the start of day $t+2$, where the same process follows and the position is reassessed. Transaction costs are assumed to be negligible for each trade that is executed.

\begin{table*}[ht] 
\centering
\caption{Out-of-sample performance for benchmark asset buy-hold strategies.}
\label{tab:passive}
\begin{tabular}{|l|c|c|c|c|c|c|}
\hline
\multicolumn{1}{|p{2.1cm}|}{\centering Benchmark \\ Asset} & \multicolumn{1}{p{1.6cm}|}{\centering Cumulative \\ Return (\%)} & \multicolumn{1}{p{1.6cm}|}{\centering Annualized \\ Expected \\ Return (\%)} & \multicolumn{1}{p{1.6cm}|}{\centering Annualized \\ Volatility (\%)} & \multicolumn{1}{p{1.6cm}|}{\centering Daily \\ Return \\ Skewness} & \multicolumn{1}{p{1.6cm}|}{\centering Daily \\ Return \\ Kurtosis} & \multicolumn{1}{p{1.6cm}|}{\centering Maximum \\ Drawdown (\%)} \\ \hline
\multicolumn{1}{|p{2.1cm}|}{\raggedright S\&P 500} & 158.63 & 8.62 & 15.67 & -0.080 & 11.958 & 27.95 \\ \hline
\multicolumn{1}{|p{2.1cm}|}{\raggedright Crude Oil} & 20.41 & -17.72 & 38.77 & -1.177 & 16.095 & 81.29 \\ 
\hline
\end{tabular}
\end{table*}

\subsection{Strategy 1: Tail-hedging}
\label{section:str1}

The tail-hedging strategy is designed for investors that seek long exposure to a given asset while being hedged against crisis periods. The strategy holds a long position during regime 1 and switches to a short position during regime 2; it is designed for assets with positive expected returns during regime 1 and negative expected returns in regime 2. We demonstrate the performance of this strategy using the continuous front month futures contracts on crude oil and the S\&P 500, respectively. 

\begin{figure}[H]
    \centering
    \includegraphics[width=\linewidth]{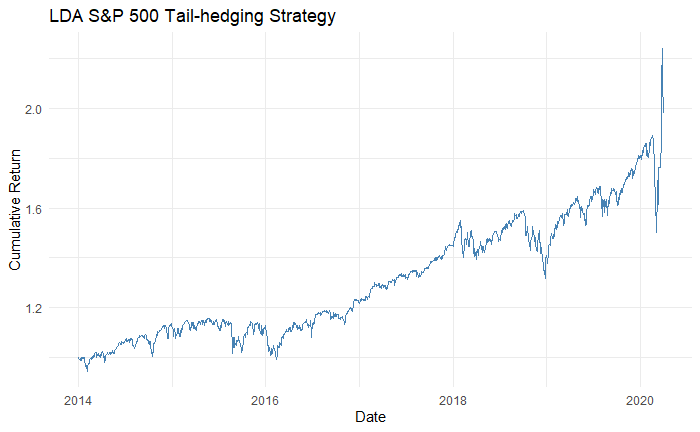}
    \caption{Out-of-sample performance of the S\&P 500 tail-hedging strategy constructed based on the LDA model's regime signals.} 
    \label{fig:str1_lda}
\end{figure}

Applied to the S\&P 500, four out of six models were able to construct a tail-hedging strategy which achieves higher cumulative returns than the benchmark; five of the models achieved a lower maximum drawdown. The strategy produces an average alpha of 5.12\% across the models while maintaining an average beta of 0.498. The cumulative return of the S\&P 500 tail-hedging strategy constructed by the LDA model can be seen in Figure~\ref{fig:str1_lda}. The strategy outperforms the benchmark S\&P 500 and is able to reduce maximum drawdown by approximately 7\%.

\begin{figure}[H]
    \centering
    \includegraphics[width=\linewidth]{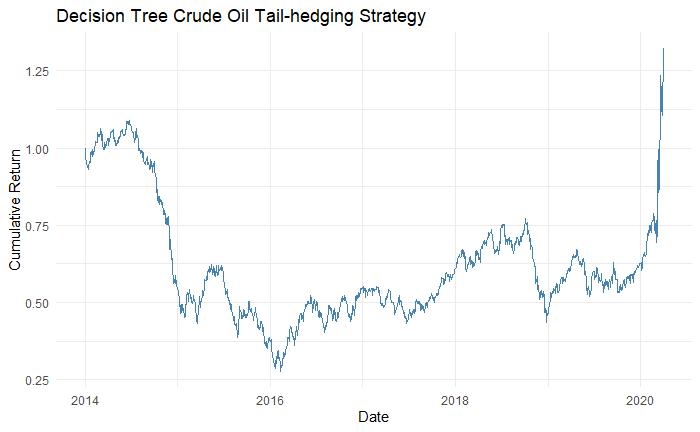}
    \caption{Out-of-sample performance of the crude oil tail-hedging strategy constructed based on the decision tree model's regime signals.} 
    \label{fig:str1_dt}
\end{figure}

When applied to crude oil, all six models achieve higher cumulative returns and lower maximum drawdowns in comparison to the passive buy-hold strategy. The tail-hedging strategy preserves an average beta of 0.523 and produces an average alpha of 13.39\% across models. Figure~\ref{fig:str1_dt} displays the cumulative return of the crude oil tail-hedging strategy constructed by the decision tree model. The strategies resulting from the decision tree and AdaBoost models exhibit nearly identical performance. The strategies realize an improved cumulative return of approximately 132\% compared to the benchmark's 20.41\% over the same period, as well as an annualized alpha of 20.53\%. The strategies also reduce maximum drawdown by 6.7\%. The performance of the tail-hedging strategy produced by each model for crude oil and the S\&P 500 is displayed in Table~\ref{tab:str1}. Despite identifying medium--strong linear relationships between in-sample performance metrics and the performance of the resulting strategies, as assessed by Pearson's correlation coefficient ($0.5<|R|$), none are statistically significant at $\alpha=.05$. As reflected in the results, the tail-hedging strategy constructed through this hybrid learning approach would allow investors to attain superior risk-adjusted returns over the out-of-sample period compared to buy and hold passive investment strategies. The regime switching based tail-hedging strategy displays the capability to effectively generate positive alpha for investors while sustaining beta exposure to underlying assets. 

\begin{table*}[ht] 
\centering
\caption{Tail-hedging strategy out-of-sample performance for the S\&P 500 and crude oil.}
\label{tab:str1}
\begin{tabular}{|cl|c|c|c|c|c|c|c|c|}
\hline
 & & \multicolumn{1}{p{1.5cm}|}{\centering Cumulative \\ Return (\%)} & \multicolumn{1}{p{1.5cm}|}{\centering Annualized \\ Expected \\ Return (\%)} & \multicolumn{1}{p{1.5cm}|}{\centering Annualized \\ Volatility (\%)} & \multicolumn{1}{p{1.5cm}|}{\centering Daily \\ Return \\ Skewness} & \multicolumn{1}{p{1.5cm}|}{\centering Daily \\ Return \\ Kurtosis} & \multicolumn{1}{p{1.5cm}|}{\centering Annualized \\ Alpha (\%)} & \multicolumn{1}{p{1.00cm}|}{\centering Beta} & \multicolumn{1}{p{1.40cm}|}{\centering Maximum \\ Drawdown (\%)}  \\ \hline
\multirow{ 6}{*}{\rotatebox[origin=c]{90}{S\&P 500}} & \multicolumn{1}{|p{2.5cm}|}{\raggedright LDA} & 202.72 & 12.54 & 15.66 & 0.054 & 11.968 & 7.02 & 0.641 & 20.63 \\ \cline{2-10}
& \multicolumn{1}{|p{2.5cm}|}{\raggedright QDA} & 164.14 & 9.16 & 15.67 & -0.112 & 11.967 & 1.79 & 0.856 & 33.50 \\ \cline{2-10}
& \multicolumn{1}{|p{2.5cm}|}{\raggedright Logistic Regression} & 158.32 & 8.59 & 15.67 & -0.809 & 12.059 & 4.78 & 0.443 & 20.64 \\ \cline{2-10}
& \multicolumn{1}{|p{2.5cm}|}{\raggedright Decision Tree} & 168.89 & 9.63 & 15.67 & -0.796 & 12.077 & 6.04 & 0.416 & 17.46 \\ \cline{2-10}
& \multicolumn{1}{|p{2.5cm}|}{\raggedright AdaBoost} & 168.89 & 9.63 & 15.67 & -0.796 & 12.077 & 6.04 & 0.416 & 17.46 \\ \cline{2-10}
&  \multicolumn{1}{|p{2.5cm}|}{\raggedright Naive Bayes} & 142.36 & 6.89 & 15.68 & -0.861 & 12.03 & 5.04 & 0.214 & 24.40 \\ \hline \hline
\multirow{ 6}{*}{\rotatebox[origin=c]{90}{Crude Oil}} & \multicolumn{1}{|p{2.5cm}|}{\raggedright LDA} & 67.22 & 1.05 & 38.78 & 1.192 & 16.196 & 10.85 & 0.553 & 74.59 \\ \cline{2-10}
& \multicolumn{1}{|p{2.5cm}|}{\raggedright QDA} & 33.29 & -9.90 & 38.78 & -1.086 & 16.142 & 5.29 & 0.857 & 75.14 \\ \cline{2-10}
& \multicolumn{1}{|p{2.5cm}|}{\raggedright Logistic Regression} & 76.59 & 3.13 & 38.78 & 1.258 & 16.179 & 12.33 & 0.519 & 74.59 \\ \cline{2-10} 
& \multicolumn{1}{|p{2.5cm}|}{\raggedright Decision Tree}  & 132.15 & 11.86 & 38.78 & 1.253 & 16.119 & 20.53 & 0.489 & 74.59 \\ \cline{2-10}
& \multicolumn{1}{|p{2.5cm}|}{\raggedright AdaBoost} & 132.15 & 11.86 & 38.78 & 1.253 & 16.119 & 20.53 & 0.489 & 74.59 \\ \cline{2-10}
&  \multicolumn{1}{|p{2.5cm}|}{\raggedright Naive Bayes} & 95.96 & 6.73 & 38.78 & 1.288 & 16.151 & 10.80 & 0.229 & 67.86 \\ 
\hline
\end{tabular}
\end{table*}

\subsection{Strategy 2: Tactical Allocation}
\label{section:str2}
\begin{figure}[h]
    \centering
    \includegraphics[width=\linewidth]{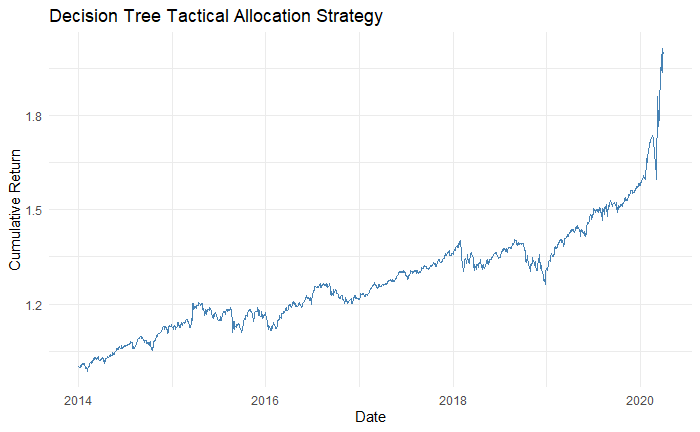}
    \caption{Out-of-sample performance of the tactical allocation strategy constructed based on the decision tree model's regime signals.} 
    \label{fig:str2_dt}
\end{figure}

Tactical allocation acts to hedge investors during crisis periods and minimize risk through multi-asset diversification, while providing investors with reasonable exposure to equity risk premium during non-crisis periods. The strategy rotates between two different portfolios. The active portfolio at a given point in time is dependent on the detected regime. Throughout regime 1, the strategy maintains a traditional 60/40 portfolio consisting of the S\&P 500 and U.S. treasury bonds. During regime 2, the strategy switches to a portfolio of short positions on the S\&P 500 and crude oil, and long positions in gold and U.S. treasury bonds, each allocated 25\% of the total portfolio weight. All six models were able to achieve higher cumulative returns and lower maximum drawdowns in comparison to the benchmark (S\&P 500). The strategy produces an average alpha of 8.24\% across all models and an average beta of approximately 0.248. Figure~\ref{fig:str2_dt} displays the simulated portfolio cumulative return for the tactical allocation strategy using the decision tree model for signal generation. Again, the portfolio performance resulting from the regime detection of both the decision tree and AdaBoost models is nearly identical, and exceeds those of the benchmark. The resulting strategies attain a significant reduction in maximum drawdown and annualized volatility with differences of 17.83\% and 5.92\%, respectively. Both strategies also show an increase of 2.91\% in annualized expected return relative to the S\&P 500. Overall, the tactical allocation strategy is able to generate positive alpha while maintaining low beta exposure to the equity market. As results show, the strategy would enable investors to capture superior risk-adjusted returns compared to the S\&P 500 benchmark over the out-of-sample period. 

The performance statistics for the tactical allocation strategy produced by each model are shown in Table~\ref{tab:str2}. The accuracy of a model, assessed by cross-validation on the in-sample data, displays a strong correlation with the cumulative return ($P=.04$), annualized expected return ($P=.04$), and maximum drawdown ($P=.01$) of the tactical allocation strategy; these correlations are $0.836$, $0.832$, and $-0.905$, all of which are statistically significant at $\alpha=.05$. Statistically significant relationships are also found between the F1 score and the aforementioned trading strategy performance statistics; the correlations are $0.836$, $0.832$, $-0.905$, while the \textit{P}--values are $P=.04$, $P=.04$, and $P=.01$, respectively. This implies that a notable association exists between these performance evaluation metrics and the out-of-sample performance of the tactical allocation strategy. In practice, one might seek a classification model which is able to maximize accuracy and F1 score in order to construct the optimal tactical allocation strategy based on regime switches. In-sample AUC shows no statistically significant relationships with any measure of performance for the tactical allocation trading strategy. 

\begin{table*}[ht] 
\centering
\caption{Tactical allocation strategy out-of-sample performance.}
\label{tab:str2}
\begin{tabular}{|l|c|c|c|c|c|c|c|c|}
\hline
& \multicolumn{1}{p{1.5cm}|}{\centering Cumulative \\ Return (\%)} & \multicolumn{1}{p{1.5cm}|}{\centering Annualized \\ Expected \\ Return (\%)} & \multicolumn{1}{p{1.5cm}|}{\centering Annualized \\ Volatility (\%)} & \multicolumn{1}{p{1.5cm}|}{\centering Daily \\ Return \\ Skewness} & \multicolumn{1}{p{1.5cm}|}{\centering Daily \\ Return \\ Kurtosis} & \multicolumn{1}{p{1.5cm}|}{\centering Annualized \\ Alpha (\%)} & \multicolumn{1}{p{1.00cm}|}{\centering Beta} & \multicolumn{1}{p{1.40cm}|}{\centering Maximum \\ Drawdown (\%)} \\ \hline
\multicolumn{1}{|p{2.5cm}|}{\raggedright LDA} & 193.55 & 11.07 & 9.96 & 2.623 & 47.864 & 8.40 & 0.309 & 11.51 \\ \hline
\multicolumn{1}{|p{2.5cm}|}{\raggedright QDA} & 167.73 & 8.75 & 9.67 & 0.159 & 18.509 & 4.92 & 0.445 & 17.59 \\ \hline
\multicolumn{1}{|p{2.5cm}|}{\raggedright Logistic Regression} & 182.09 & 10.06 & 9.69 & 2.669 & 50.799 & 8.13 & 0.224 & 11.51 \\ \hline 
\multicolumn{1}{|p{2.5cm}|}{\raggedright Decision Tree} & 199.46 & 11.53 & 9.75 & 2.635 & 49.385 & 9.73 & 0.209 & 10.12 \\ \hline
\multicolumn{1}{|p{2.5cm}|}{\raggedright AdaBoost} & 199.46 & 11.53 & 9.75 & 2.635 & 49.385 & 9.73 & 0.209 & 10.12 \\ \hline
\multicolumn{1}{|p{2.5cm}|}{\raggedright Naive Bayes} & 173.45 & 9.35 & 10.37 & 1.933 & 38.015 & 8.55 & 0.093 & 14.96 \\
\hline
\end{tabular}
\end{table*}

\section{Conclusion and Future Works}
\label{section:conclusion}
In this work, we present a novel approach for the detection of regime switches using a combination of unsupervised and supervised learning techniques. Specifically, we show that cluster analysis with classification provides a hybrid framework for regime detection. We identify and exhibit one method for inferring and subsequently automating the selection of the number of regimes which exist through the average silhouette width method. Further, we demonstrate a practical application of this approach through the construction of two successful trading strategies in the financial markets; though, specific use-cases may be adopted within areas of portfolio management and systematic trading. 

Building on the use of economic data in this study, assessing the effectiveness of this approach using asset specific factors and statistics may be a natural extension of this work. In this study we use differenced time-series which lack the memory that is often intrinsically embedded within time-series data -- other pre-processing steps may be worth exploring in order to maximize information retention. Due to the approach being model agnostic it is desirable to consider the use of other measures of distance and dissimilarity, as well as clustering methods such as those which are hierarchical or density-based. Assessing how these methods influence the number and types of regimes inferred from data could allow for the development of systems configured specifically for collections of related time-series. Although associations are found between in-sample performance metrics and out-of-sample trading strategy performance in this study, the significance of such relationships is not directly evident. Further study is necessary to better understand these relationships, and would increase the confidence in the out-of-sample performance produced by trading strategies resultant from the proposed hybrid framework.

\begin{acks}
We would like to express a deep gratitude to Professor Reinaldo Sanchez-Arias at Florida Polytechnic University and Professor Jiandong Ren at Western University for valuable suggestions in the development of this work. We also wish to acknowledge Shivam Sharma, as well as colleagues Neha Gulati and Carol Zhai for the constructive comments and useful critiques of this research.
\end{acks}

\bibliographystyle{ACM-Reference-Format}
\bibliography{ICAIF2020}

\end{document}